\documentclass[reprint,aps, longbibliography, groupedaddress]{revtex4-1}
\usepackage{graphicx,amsmath,amssymb,braket,siunitx,mathtools}
\usepackage[dvipsnames]{xcolor}
\usepackage[unicode]{hyperref}
\DeclareSIUnit\gauss{G}

\DeclarePairedDelimiterXPP\BigOSI[2]%
  {\mathcal{O}}{(}{)}{}%
  {\SI{#1}{#2}}

\hypersetup{
    colorlinks=true,
    citecolor=Blue,
    linkcolor=Blue,
    urlcolor=Blue
}
\newcommand{\mref}[2]{%
    \hyperref[{#1}]{%
        \ref*{#1}#2%
    }%
}
\begin{document}
\title{Enhancing dipolar interactions between molecules using state-dependent optical tweezer traps}
\author{L. Caldwell}
\author{M. R. Tarbutt}
\affiliation{Centre for Cold Matter, Blackett Laboratory, Imperial College London, Prince Consort Road, London SW7 2AZ UK
}

\begin{abstract}
    We show how state-dependent optical potentials can be used to trap a pair of molecules in different internal states at a separation much smaller than the wavelength of the trapping light. 
    This close spacing greatly enhances the dipole-dipole interaction and we show how it can be used to implement two-qubit gates between molecules that are 100 times faster than existing protocols and than rotational coherence times already demonstrated.
    We analyze complications due to hyperfine structure, tensor light shifts, photon scattering and collisional loss, and conclude that none is a barrier to implementing the scheme.
\end{abstract}

\maketitle

Electric dipole-dipole interactions can be used to entangle polar molecules. Ensembles of such molecules are a promising platform for quantum simulation \cite{Barnett2006,Buchler2007, Micheli2007}, quantum computation \cite{DeMille2002} or the creation of many-body states for precision metrology \cite{Hazzard2013}. Recent progress in the production and control of ultracold molecules \cite{Danzl2008,Ni2008,Barry2014,Prehn2016,Truppe2017b, Collopy2018,Cheuk2018, Williams2018,DeMarco2019,Caldwell2019} has brought these goals within reach of near-term experiments.

Molecules confined in arrays of optical tweezer traps are particularly attractive and have recently been realized~\cite{Liu2018, Liu2019, Anderegg2019, Zhang2020}. The platform is scalable to several hundred sites, enables re-arrangement of the traps \cite{Endres2016, Barredo2018} to reduce entropy or control which particles interact, and provides natural single-site addressability.
Various authors have proposed protocols for two-qubit gates using rotational states of molecules \cite{DeMille2002, Andre2006, Yelin2006,Pellegrini2011, Ni2018,Hudson2018,Hughes2020}. The number of possible gate operations is set by the ratio $E_{\rm dd}\tau_c/h$ where $E_{\rm dd}$ is the dipole-dipole interaction energy and $\tau_c$ is the coherence time of a trapped molecule in a superposition of rotational states. For conventional tweezer traps $E_{\rm dd}$ is limited by the minimum trap separation. This is roughly the wavelength of the trapping light, typically $\sim \SI{1}{\micro\meter}$, giving $E_{\rm dd}/h\sim \SI{1}{\kilo\hertz}$.
Recent work has extended $\tau_c$ to several milliseconds \cite{Seebelberg2018, Caldwell2020} but, at these interaction strengths, only a few high-fidelity gates are possible. While prospects are good for further improvements---coherence times near 1~s have been demonstrated in hyperfine states of molecules \cite{Park2017} and electronic states of atoms in tweezers \cite{Norcia2019}---considerable advances are required to realize the full potential of this platform for quantum science.

\begin{figure}
    \centering
    \includegraphics{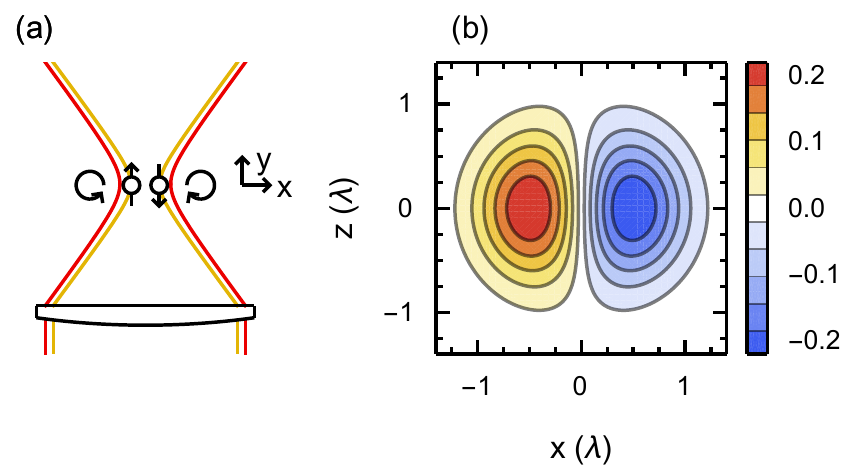}
    \caption{(a) State-dependent trap formed from a two-color optical tweezer. Two molecules in different internal states are trapped in different locations due to opposite circular handedness (rotating arrows) on opposite sides of the focus. (b) Contour plot of $(I/I_{\rm max})\vec{C}\cdot\hat{z}$ in $y=0$ focal plane for a single tweezer. Calculated using the vector Debye integral \cite{Richards1959} for a lens with ${\rm NA}=0.55$. The input beam is polarized along $x$ and has $1/e^2$ diameter equal to that of the lens. An approximate analytical approach to this calculation, based on Ref.~\cite{*[{See, for example, Section 3.3 of }] [{.}] Novotny2012}, is given in the SM.}
    \label{fig:polstruc}
\end{figure}

Here we show how to increase the dipole-dipole interaction between two molecules by trapping them at reduced separations using the state-dependence of the molecule-light interaction. Our scheme has similarities to state-dependent optical lattices which have been used to control atoms on sub-wavelength scales \cite{Brennen1999, Jaksch1999,Mandel2003,Mandel2003b,Anderlini2007, Daley2008, Karski2009, Gadway2010, Riegger2018}, but benefits from the advantages of the tweezer platform noted above.
For atoms, electric dipole-dipole interactions involve electronically excited states \cite{Brennen1999} whose short lifetimes severely limit $\tau_c$. By using long-lived rotational states of molecules, we avoid this limitation entirely.
Our method, shown in Fig.~\mref{fig:polstruc}{(a)}, uses two optical tweezers of different wavelengths, focused at the same position. The tight focussing of the light produces elliptical polarization components with opposite handedness on each side of the focus \cite{Thompson2013, Wang2019}. A molecule with non-zero spin has an interaction with the light field that depends both on this handedness and on the orientation of the spin. Consequently, two molecules in different internal states are trapped at different positions in the trap and their separation can be controlled by varying the relative intensities of the two tweezers. This state-dependent potential allows $E_{\rm dd}$ to be enhanced by two orders of magnitude. We introduce these concepts and show how to apply them in practice to implement fast two-qubit gates.

\begin{figure*}
    \centering
    \includegraphics{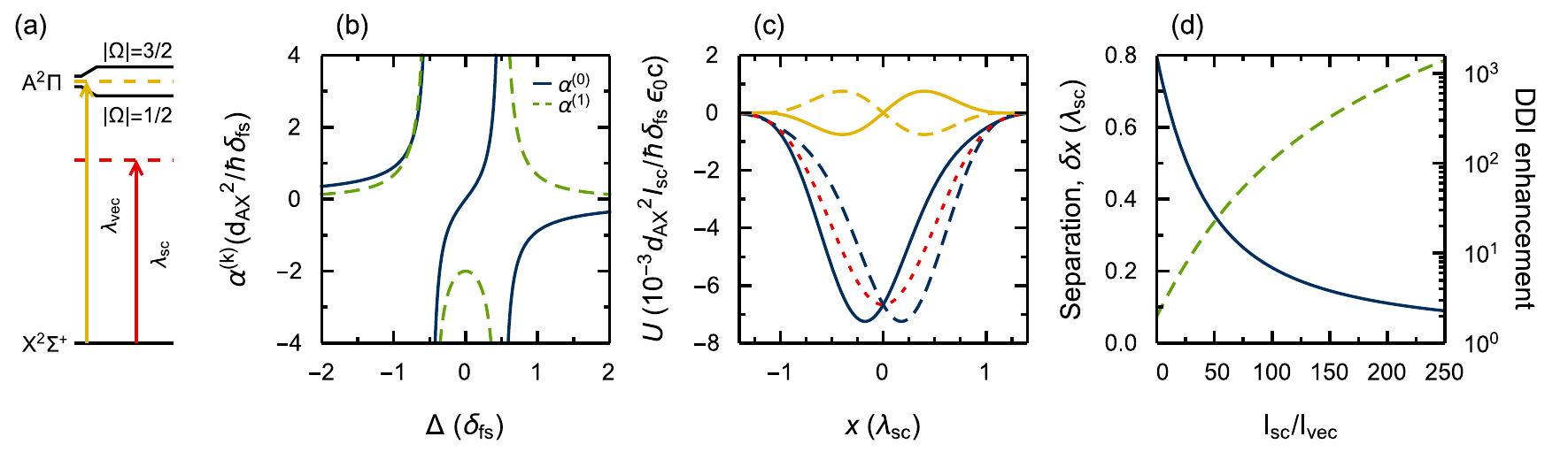}
    \caption{(a) Schematic of relevant electronic structure. The excited state $A^2\Pi$ is split by spin-orbit coupling into two states with $|\Omega|=\tfrac{1}{2},\tfrac{3}{2}$. Dashed lines: energies of scalar (red) and vector (orange) tweezers. (b) Scalar (blue) and vector (green dashed) polarizabilities calculated using Eqs.~\eqref{eq:polarisabilities}. (c) Calculated potentials of scalar (red dotted), vector (orange) and combined (blue) tweezers versus displacement along $x$ for $J=\tfrac{1}{2}$ states of $N=0$ and 1. Solid (dashed) line shows the potential for  $m_J=-\tfrac{1}{2}(\tfrac{1}{2})$. Calculations are for $\lambda_{\rm vec}=0.8\lambda_{\rm sc}$, $\Delta_{\rm sc}=50\delta_{\rm fs}$ and $I_{\rm sc}/I_{\rm vec}=50$. (d) Blue line, left axis: $\delta x$ versus $I_{\rm sc}/I_{\rm vec}$. Green dashed line, right axis: enhancement of $E_{\rm dd}$ for point particles trapped at the two minima.}
    \label{fig:potentials}
\end{figure*}

\textit{Method.}---The scheme can be illustrated using a simple $^2\Sigma$ molecule with no hyperfine structure. We focus on the four states with total angular momentum $J=\tfrac{1}{2}$,
\begin{equation}
\begin{split}
    \ket{0_\pm} &= \ket{N=0,J=\tfrac{1}{2},m_J=\pm\tfrac{1}{2}},\\
    \ket{1_\pm} &= \ket{N=1,J=\tfrac{1}{2},m_J=\pm\tfrac{1}{2}},\label{eq:model-mol-states}
\end{split}
\end{equation}
where the pair of states with rotational angular momentum $N=0$ are separated from the pair with $N=1$ by the rotational energy $E_{\rm rot}$.

Consider the interaction of this molecule with a light field of intensity $I$ and polarization $\vec{\epsilon}$. The interaction has scalar, vector and tensor parts whose dependence on the frequency of the light can be factored out into three constants $\alpha^{(0)}$, $\alpha^{(1)}$ and $\alpha^{(2)}$; the scalar, vector and tensor polarizabilities.
The scalar interaction shifts all four of our states by $W_0 = -\frac{1}{2 \epsilon_0 c}\alpha^{(0)} I$.
The vector and tensor parts cause state-dependent shifts.
The vector shift is non-zero when the field has ellipticity, described by $\vec{C}=\mathrm{Im}(\vec{\epsilon}\times\vec{\epsilon}^{\,*})$. $|\vec{C}|$ gives the degree of ellipticity and its direction gives the handedness. For incident light propagating along $y$ and linearly polarized along $x$, this handedness is along $z$ and is opposite either side of the focus (see Fig.~\mref{fig:polstruc}{(b)} and Supplemental Material (SM)~\footnote{See Supplemental Material for further details of polarization structure, polarizability, the dipole-dipole interaction, photon scattering and collisional loss, and loading of the state-dependent traps.}). In this case the vector shift is $W_1 = \frac{1}{2 \epsilon_0 c}\alpha^{(1)} g_J m_J (\vec{C}\cdot\hat{z}) I$, where $g_J=1/[2J(J+1)]$ and we have assumed $W_1$ is small compared to the spin-rotation interaction. $W_1$ is identical for $\ket{0_-/1_-}$ and opposite to that of $\ket{0_+/1_+}$.
The tensor shift is zero for our $J=\tfrac{1}{2}$ states; we return to it later.

The polarizabilities $\alpha^{(k)}$ depend on the details of the electronic structure~\cite{Caldwell2020b}. Here, for simplicity, we assume that they are dominated by interaction with the first excited electronic state. The relevant electronic structure is shown in Fig.~\mref{fig:potentials}{(a)}. The spin-orbit interaction splits the excited state into two components separated by $\delta_{\rm fs}$, typically $\sim2\pi\times \SI{1}{\tera\hertz}$. Their mid-point is $\omega_{AX}$ above the ground state, typically $\sim2\pi \times\SI{500}{\tera\hertz}$, and we define $\Delta$ as the detuning of the light field from this point. For $\Delta \ll \omega_{AX}$, the polarizabilities can be written (see SM)
\begin{equation}
    \begin{split}
    \alpha^{(0)} \simeq -\frac{4\Delta}{3 \delta_{\rm fs}} \alpha^{(1)} \simeq - \alpha^{(2)} \simeq \frac{-2 d_{AX}^2 \Delta}{3\hbar(\Delta-\frac{\delta_{\rm fs}}{2})(\Delta+\frac{\delta_{\rm fs}}{2})},\label{eq:polarisabilities}
\end{split}
\end{equation}
where $d_{AX}$ is the dipole matrix element connecting the $X^2\Sigma$ and $A^2\Pi$ states.

We use tweezer traps at two different wavelengths $\lambda_{\rm sc}$ and $\lambda_{\rm vec}$, shown schematically in Fig.~\mref{fig:potentials}{(a)}, which we call the scalar and vector traps. Their on-axis intensities are $I_{\rm sc}$ and $I_{\rm vec}$. The scalar trap light is red-detuned with $\Delta=\Delta_{\rm sc}\gg\delta_{\rm fs}$. In this regime, $\alpha^{(0)} \gg \alpha^{(1)}$ and the interaction is dominated by the scalar component. The vector trap light is tuned between the fine structure components. Figure~\mref{fig:potentials}{(b)} shows $\alpha^{(0)}$ and $\alpha^{(1)}$ in this region. When $\Delta=\Delta_{\rm vec}=0$, $\alpha^{(0)}=0$ while $\alpha^{(1)}=-2 d_{AX}^2/\hbar\delta_{\rm fs}$ which can be large.

Figure~\mref{fig:potentials}{(c)} shows the light shifts of the four $J=\tfrac{1}{2}$ states of our model molecule as a function of position along the $x$ axis. The dotted red line shows the light shift of all four states in the scalar trap for $\Delta_{\rm sc}=50\delta_{\rm fs}$. The tiny state dependence caused by the residual vector light shift at $\lambda_{\rm sc}$ is not visible on this scale. The solid (dashed) orange line shows the light shift of the $\ket{0_-/1_-}$ ($\ket{0_+/1_+}$) states in the vector trap where we have assumed $\lambda_{\rm vec}=0.8\lambda_{\rm sc}$ and $I_{\rm sc}/I_{\rm vec}=50$. Molecules in the $\ket{0_-/1_-}$ states and the $\ket{0_+/1_+}$ states are trapped on opposite sides of the focus.
When both scalar and vector traps are present the potentials add, shown by the blue lines. Varying their relative intensity controls the separation of the minima, $\delta x$. The blue line in Fig.~\mref{fig:potentials}{(d)} shows how $\delta x$ depends on the intensity ratio while the green line shows the enhancement of $E_{\rm dd}$ for two point particles positioned at the trap minima relative to those in separated scalar traps. The enhancement is ultimately limited by undesirable collisions that occur when the spacing is too small. As we will see, for realistic parameters, an enhancement of 2 orders of magnitude is achievable.

\textit{Eigenstates.}---The dipole-dipole interaction Hamiltonian is
\begin{equation}
    \begin{split}
    H_{\rm dd}=&\frac{\vec{d}_A\cdot \vec{d}_B - 3 (\vec{d}_A\cdot\hat{x})\otimes(\vec{d}_B\cdot\hat{x})}{4\pi\epsilon_0 |x_B-x_A|^3},\label{eq:dipole-dipole}
    \end{split}
\end{equation}
where $\vec{d}_A$ and $\vec{d}_B$ are the dipole moments of the two molecules, $x_A$ and $x_B$ their positions, and $\hat{x}$ is a unit vector along $x$. As we show in the SM, after restricting ourselves to states with one molecule trapped on either side of the focus, the eigenstates of the two-molecule Hamiltonian, including $H_{\rm dd}$, are
\begin{equation}
    \begin{split}
        \ket{00} = \ket{0_-}\ket{0_+}, \quad \ket{11} = \ket{1_-}\ket{1_+},\\
        \ket{\Psi^\pm} = \frac{1}{\sqrt{2}}\left(\ket{0_-}\ket{1_+} \pm \ket{1_-}\ket{0_+}\right),\label{eq:two-mol-states}
    \end{split}
\end{equation}
with energies $0$, $2 E_{\rm rot}$ and $E_{\rm rot}\pm E_{\rm dd}$ respectively. Here $E_{\rm dd}=\Lambda_{10}/4\pi\epsilon_0|x_B-x_A|^3$ and the quantity $\Lambda_{ij}=\bra{j_-}\bra{i_+}\vec{d}_A\cdot \vec{d}_B - 3 (\vec{d}_A\cdot\hat{x})\otimes(\vec{d}_B\cdot\hat{x})\ket{i_-}\ket{j_+}$ can be positive or negative.

Since $|E_{\rm dd}|$ can approach or even exceed the motional energy spacing in the trap, $\hbar \omega_{\rm t}$, it is important to consider the motional degree of freedom of the two molecules. A 1D treatment is sufficient to elucidate the main points. When the upper and lower states in each pair have the same vector shift, so that the potential is the same for both, the eigenstates are (see SM) $\ket{\psi}\ket{n_{\rm cm}}\ket{\phi(x_{\rm rel})}$. Here $\ket{\psi}$ is one of the internal eigenstates of Eqs.~\eqref{eq:two-mol-states} and $\ket{n_{\rm cm}}$ is a harmonic oscillator eigenstate for the center of mass coordinate $x_B+x_A$. The relative motional state $\ket{\phi(x_{\rm rel})}$ is an eigenstate of the state-dependent dimensionless Hamiltonian
\begin{equation}
    H_{\rm rel} = \frac{p_{\rm rel}^2}{2} +\frac{1}{2}(x_{\rm rel} - \tilde{\delta x})^2 + q\frac{r^3}{|x_{\rm rel}|^3}.\label{eq:Hpm}
\end{equation}
Here $x_{\rm rel}=\sqrt{\frac{M\omega_{\rm t}}{2\hbar}}(x_B-x_A)$ is the reduced relative motional coordinate, $p_{\rm rel}$ the conjugate momentum, $\tilde{\delta x}=\sqrt{\frac{M\omega_{\rm t}}{2\hbar}}\delta x$, $r = \sqrt{\frac{M\omega_{\rm t}}{2\hbar}}(\Lambda_{10}/4\pi\epsilon_0\hbar\omega_{\rm t})^{1/3}$ is the separation, in reduced units, at which $E_{\rm dd} = \hbar \omega_{\rm t}$, and $M$ is the mass of the molecule. The factor $q$ reflects the state-dependence of the dipole-dipole interaction, and is equal to $\{0,-1,1,0\}$ for $\ket{\psi}=\{\ket{00},\ket{\Psi^-},\ket{\Psi^+},\ket{11}\}$ respectively.
The relative motional states for $q r^3 >0$ are examined in the SM and show two important effects. First, the finite extent of the wavefunction means that $\langle 1/x_{\rm rel}^3 \rangle > 1/\langle x_{\rm rel} \rangle^3$. Second, the molecules are pushed apart by their interaction so their mean separation is larger than $\delta x$. 
In the motional ground state, the first effect dominates at larger $\delta x$ increasing $E_{\rm dd}$, while the second effect dominates at small $\delta x$, reducing $E_{\rm dd}$ below the value for fixed point dipoles.

\textit{Complications in real molecules.}---The addition of nuclear spin introduces a hyperfine interaction which can mix states of different $J$. For these mixed states, the vector Stark shift depends on the relative size of the hyperfine and spin-rotation interactions, which differs from one rotational state to the next. Consequently, the position of the potential minimum for $\ket{0_{\pm}}$ is shifted relative to $\ket{1_{\pm}}$. As shown in the SM, for a shift $\xi$ in reduced units, the resulting imperfect overlap of the spatial wavefunctions reduces the dipole-dipole energy by $e^{-\xi^2}$, the square of the overlap integral. As we will see, this reduction is typically small. 

A second complication is that states outside $N=0$ can have a tensor Stark shift due to the light at $\lambda_{\rm sc}$. The most relevant effect of this is to couple states with $\Delta m \le 2$ near the center of the trap, allowing tunneling between the left and right potentials. This coupling is eliminated when the incident polarization of the scalar and vector traps are orthogonal. At other angles the tunneling is proportional to the wavefunction overlap so becomes negligible when the molecules are well separated. 

\begin{figure}
    \centering
    \includegraphics{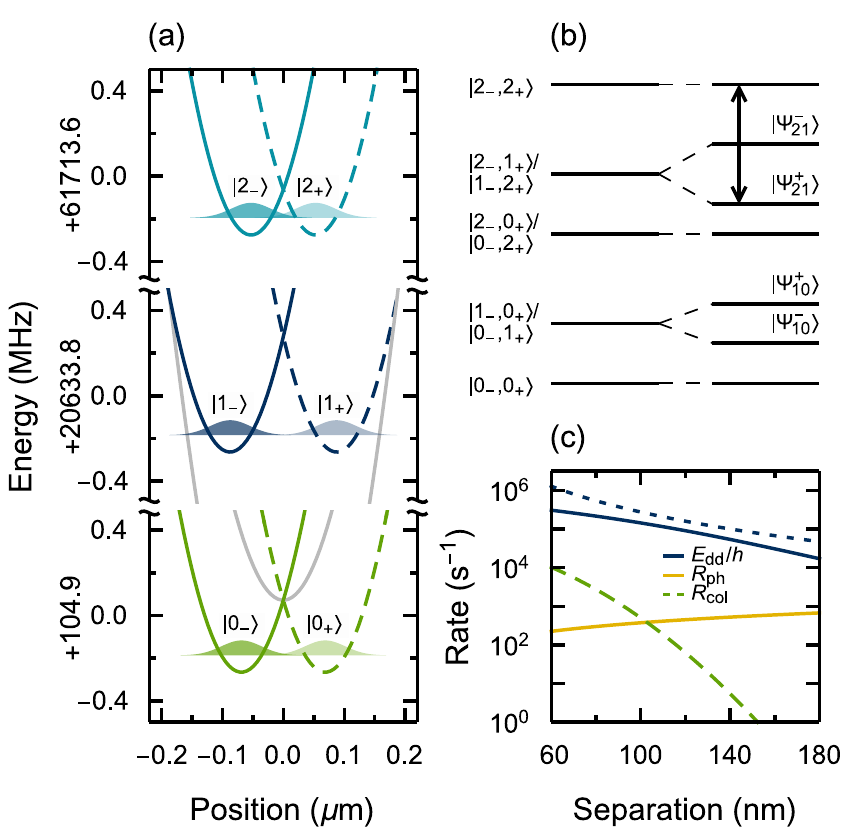}
    \caption{Two-qubit gate in CaF. (a) Lines: potentials for the states used, calculated for \SI{10}{\milli\watt} \SI{780}{\nano\meter} scalar tweezer with incident polarization along $z$ and \SI{35.4}{\micro\watt} \SI{604.966}{\nano\meter} vector tweezer polarized along $x$. The lens has NA of 0.55. Input beams have $1/e^2$ diameter equal to the lens diameter. Wavepackets show ground motional states of each potential. Energies are relative to the $\ket{N=0,F=0}$ ground state in zero field. (b) Level diagram of two-molecule states relevant to the gate, not to scale. Left (right) half shows energies of states without (with) dipole-dipole interaction. Black arrow: transition driven for two-qubit gate. (c) Dipole-dipole energy, photon scattering rate and collisional loss rate versus separation of $\ket{2_\pm}$ states. Dotted blue line: energy for fixed point dipoles; solid line: full 1D calculation.}
    \label{fig:gate}
\end{figure}

\textit{Realistic example.}---To illustrate the power and practicality of our method, we show how to implement a simple two-qubit gate using CaF molecules. CaF has  been confined in optical tweezer traps \cite{Anderegg2019} and has a structure similar to the model molecule but with a fluorine nuclear spin of $\tfrac{1}{2}$. 
We map the states of the model molecule to our specific case as follows: $\ket{0_\pm}=\ket{N=0,F=1,m_F=\pm1}$, $\ket{1_\pm}=\ket{1,1,\pm1}$. We also introduce $\ket{2_\pm}=\ket{2,2,\pm2}$ \footnote{Note that there are two levels with $F=1$ in $N=1$ and two with $F=2$ in $N=2$. In both cases we choose the higher lying level which has a larger vector Stark shift.}. The states $\ket{0_\pm}$ and $\ket{2_\pm}$ form our computational basis, while $\ket{1_\pm}$ are used to implement the gate. Many other choices of states are possible and may have different advantages. Figure~\mref{fig:gate}{(a)} shows the potentials for the three pairs of states and for the parameters given in the caption. The closely spaced traps can be loaded adiabatically and without collisional loss from two separated tweezers using simple intensity ramps as shown in the SM.
We suppose the molecules have been cooled to the motional ground state~\cite{Caldwell2020b}, resulting in the wavefunctions shown for each potential.
The trap frequency is within 2\% of $ \SI{160}{\kilo\hertz}$ for all states, and the corresponding rms wavepacket size is \SI{23}{\nano\meter}.

Figure~\mref{fig:gate}{(b)} shows the two-molecule states relevant for the gate. The matrix elements of $H_{\rm dd}$ are zero between states of our computational basis $\ket{0_\pm}$ and $\ket{2_\pm}$. 
The states $\ket{2_-,1_+}$ and $\ket{1_-,2_+}$ are mixed by $H_{\rm dd}$ giving the pair of entangled states $\ket{\Psi_{21}^\pm}=(\ket{2_-,1_+}\pm\ket{1_-,2_+})/\sqrt{2}$, split by $2 E_{\rm dd}$. 
A microwave pulse resonant with the $\ket{2_-,2_+}\leftrightarrow\ket{\Psi_{21}^+}$ transition and of sufficient duration to resolve it from $\ket{2_-,2_+}\leftrightarrow\ket{\Psi_{21}^-}$ entangles the two molecules. As we will see, $E_{\rm dd}$ and $\hbar\omega_{\rm t}$ are similar in size for our choice of parameters so a gate which resolves the dipole-dipole splitting will also resolve the motional sidebands so that motional heating can be avoided.
Note that the transitions $\ket{2_-,0_+}\leftrightarrow\ket{\Psi_{10}^-}$ and $\ket{0_-,2_+}\leftrightarrow\ket{\Psi_{10}^-}$ are detuned because $\Lambda_{21}/\Lambda_{10}\simeq 2.9$.
A $2\pi$ pulse implements the two-qubit gate $a\ket{0_-,0_+}+b\ket{0_-,2_+}+c\ket{2_-,0_+}+d\ket{2_-,2_+} \rightarrow a\ket{0_-,0_+}+b\ket{0_-,2_+}+c\ket{2_-,0_+}-d\ket{2_-,2_+}$.
This gate is universal in combination with single-qubit operations which can be carried out rapidly using two-photon microwave pulses \cite{Williams2018, Blackmore2018, Caldwell2020, Cheuk2020}. In an array of such qubits, single-qubit addressability is obtained through a combination of microwave polarization and tweezer intensity. The polarization determines which molecule in a pair is addressed, and a small change in intensity of the selected tweezer relative to all others ensures that only the molecule in that tweezer is addressed. 

The blue lines in Fig.~\mref{fig:gate}{(c)} show the energy shift of $\ket{\Psi_{21}^+}$ as a function of the separation of the potential minima for the $\ket{2_{\pm}}$ states. The dashed and solid lines show results for fixed point dipoles and the full 1D calculation respectively (see SM for details). For a separation of \SI{106}{\nano\meter}, as shown in Fig.~\mref{fig:gate}{(a)}, the combined effect of the dipole-dipole interaction pushing the molecules apart and the imperfect overlap of the motional wavefunctions, reduces $E_{\rm dd}$ by $\sim 45\%$. Also shown are the expected dominant loss mechanisms in the trap. We calculate the collisional loss rate $R_{\rm col}$ using the coefficient measured in Ref.~\cite{Cheuk2020} for CaF in a \SI{780}{\nano\meter} tweezer trap. $R_{\rm col}$ decreases with increasing separation and is largest when both molecules are in $\ket{2_\pm}$ where their overlap is largest. The photon scattering rate $R_{\rm ph}$ is dominated by scattering from the vector trapping light. We have assumed a fixed $I_{\rm sc}$ so $R_{\rm ph}$ increases with separation since larger separations require  larger $I_{\rm vec}$. Over the range shown, the ratio of $E_{\rm dd}/h$ to the sum of the loss rates is large. Choosing the separation of the $\ket{2_{\pm}}$ states to be \SI{106}{\nano\meter}, $R_{\rm col}\simeq \SI{280}{\per\second}$ and $R_{\rm ph}\simeq \SI{400}{\per\second}$ while $E_{\rm dd}/h=\SI{130}{\kilo\hertz}$, more than 300 times larger. This is also 100 times larger than the maximum interaction energy achievable with separate tweezers. For a fixed vector Stark shift, $R_{\rm ph}$ scales inversely with the fine-structure interval, so will be smaller for heavier molecules. For example, it is reduced by factors of $\sim 4$, 6 and 19 in SrF, YO and YbF respectively. $R_{\rm col}$ may be very different in other systems or for the same system at different wavelengths \cite{Christianen2019, Gregory2020}; this is an important topic for investigation.

To scale our scheme to many molecules, traps can be rearranged to implement gates between different pairs.  A useful metric is the time required to move a pair from two separated potentials into a single, combined trap ready for the fast gate. In the SM we describe a simple adiabatic protocol which takes \SI{50}{\micro\second}. More sophisticated non-adiabatic transport protocols \cite{Guery-Odelin2014, Couvert2008} can be completed more rapidly and without heating, as demonstrated for ions \cite{Walther2012}.

\textit{Summary.}---We have proposed a new scheme which uses state-dependent optical tweezer traps to confine pairs of polar molecules at distances much smaller than the wavelength of the trapping light, and shown how to engineer a greatly enhanced dipole-dipole interaction between them. We have analyzed an example in detail, including the effects of hyperfine structure and tensor light shifts. We find that two-qubit gates can be implemented at least 100 times faster than existing protocols with characteristic figure of merit $E_{\rm dd}\tau_{\rm c}/h \sim 10^3$ for rotational coherence times already demonstrated for molecules \cite{Seebelberg2018, Caldwell2020}. Thus, our work enables useful quantum information processing without further improvements to coherence times. Because the gate is so much faster, the effects of fluctuating magnetic fields or tweezer intensity matter less. We have designed a specific two-qubit gate, but our scheme provides a similar speedup for any gate that uses the dipole-dipole interaction e.g.~\cite{Ni2018}. Shaped microwave pulses that produce remarkable robustness to various experimental imperfections \cite{Hughes2020} can also be utilized in our scheme. Our method will work for all molecules laser cooled so far, and the heavier ones have a reduced scattering rate which may be an important advantage. The method should also work for heteronuclear bialkali molecules prepared in the $^3\Sigma$ state \cite{Rvachov2017}.

As well as quantum information processing, the enhanced dipole-dipole interactions will be useful in quantum simulation. For example, a linear chain of tweezers with a pair of molecules in each can implement an SSH model~\cite{Su1979} in a natural way. Furthermore, the ability to control the wavefunction overlap between two molecules with such precision is unique and offers a new tool for studying collisions and quantum chemistry with unprecedented precision and control.

We thank Jeremy Hutson, Jordi Mur-Petit, Paolo Molignini, Simon Cornish, Michael Hughes and Alex Guttridge for helpful discussions and feedback. This work was supported by EPSRC grant EP/P01058X/1.
\bibliography{references}

\end{document}


\title{Enhancing dipolar interactions between molecules using state-dependent optical tweezer traps: Supplemental Material}

\author{L. Caldwell}
\author{M. R. Tarbutt}
\affiliation{Centre for Cold Matter, Blackett Laboratory, Imperial College
London, Prince Consort Road, London SW7 2AZ UK}


\maketitle

\section{Polarization gradients in an optical tweezer}

Throughout, we use the vector Debye integral \cite{Richards1959} to calculate the electric field of the trapping light as a function of position. An approximate analytical approach can also be used as follows. In the focal plane, and using the coordinates of Fig.~\ref{fig:polstruc}, the longitudinal component of the electric field is related to the transverse component by
\begin{equation}
    E_y(x,0,z)\simeq -i\frac{2x}{k w_0^2} E_x(x,0,z),
\end{equation}
where $k$ is the wavevector of the trapping light and $w_0$ is the beam-waist radius \cite{Novotny2012}. The vector $\vec{C}$ is then
\begin{equation}
    \vec{C} \simeq \frac{4 k w_0^2 x}{k^2 w_0^4 + 4 x^2}\hat{z}.
\end{equation}

\section{ac Stark shift}

We calculate the ac Stark shift of a $^{2}\Sigma$ molecule following the Appendix of Ref.~\cite{Caldwell2020b}. In this section, we refer to equation numbers from that reference. The interaction of the molecule with light of polarization $\vec{\epsilon}$ and electric field magnitude $\mathcal{E}_0$ is described by the operator
\begin{equation}
H_{\rm S}=-\frac{\mathcal{E}_0^2}{4}\sum _{K=0}^2 \sum _{P=-K}^K  (-1)^P {\mathcal A}^{K}_P \mathcal{P}_{-P}^K,
\label{EqApp:HStark}
\end{equation}
where ${\mathcal A}^{K}$ are the polarizability operators and $\mathcal{P}^K$ are the polarization tensors of the light. Here, we focus on the $K=1$ term which gives the vector Stark shift, $W_1$. For incident light polarized along $x$ and propagating along $y$, $\epsilon_{0} = 0$ everywhere, and the only non-zero component of $\mathcal{P}^1$ is $\mathcal{P}_0^1 = \vec{C}\cdot\hat{z} = \textrm{Im}(\vec{\epsilon} \times \vec{\epsilon}\,^*)_0 = |\epsilon_1|^2 - |\epsilon_{-1}|^2$. Thus, we need only calculate the matrix elements of ${\mathcal A}^{1}_0$. They are diagonal in the magnetic quantum number $m_J$. Furthermore, if we specialize to the case where the Stark shift is small compared to the spin-rotation interaction, we see that $W_1=-\frac{1}{4}\mathcal{E}_0^2 \langle {\mathcal A}^{1}_0 \rangle \mathcal{P}_0^1$. Using Eq.~(A25) of Ref.~\cite{Caldwell2020b}, we find $\langle {\mathcal A}^{1}_0 \rangle = -\alpha^{(1)} m_J/(2J(J+1))$. So we reach the result
\begin{equation}
    W_1 = \frac{1}{2\epsilon_0 c} \alpha^{(1)}\frac{m_J}{2J(J+1)} (\vec{C}\cdot\hat{z}) I,
\end{equation}
where $I$ is the intensity of the light.

The expressions for the polarizability components $\alpha^{(K)}$ are given by Eqs.~(A24) and (A26) and require the evaluation of a sum over all excited states. In the main text, to illustrate our scheme, we focus on the case where only the $A^2\Pi$ excited state contributes to the polarizability. This is reasonable provided the detuning of the light field from this state is small compared to the detuning from higher-lying states. Including the effects of higher-lying electronic states changes the quantitative details but the qualitative features important to our scheme are unchanged. If the $|\Omega|=\tfrac{1}{2}$ and $\tfrac{3}{2}$ components of this state are $\hbar \omega_{1/2}$ and $\hbar \omega_{3/2}$ above the ground state respectively, we can define the fine structure splitting $\delta_{\rm fs}=\omega_{3/2}-\omega_{1/2}$ and $\Delta$, the detuning of the laser field from their midpoint at $\omega_{AX}=\tfrac{1}{2}(\omega_{3/2}+\omega_{1/2})$. In the limit where $\Delta \ll \omega_{AX}$, the expressions from Ref.~\cite{Caldwell2020b} reduce to those of Eqs.~\eqref{eq:polarisabilities} in the main text.

For our realistic example of CaF in the main text, we estimate the polarizability components from data on the $A^{2}\Pi\leftrightarrow X^{2}\Sigma$ and $B^{2}\Sigma\leftrightarrow X^{2}\Sigma$ transitions at our chosen wavelengths of $\lambda_{\rm sc}=\SI{780}{\nano\meter}$, $\lambda_{\rm vec}=\SI{604.966}{\nano\meter}$. The scalar, vector and tensor Stark shifts of all the relevant levels are calculated using the matrix elements given by Eq. (A25) of Ref.~\cite{Caldwell2020b}.

\section{Dipole-dipole interaction}

Consider the two-molecule states $\ket{F_A,m_A}\ket{F_B,m_B}$ where $m_{A/B}$ are the magnetic quantum numbers of molecules A and B and $F_{A/B}$ stand for all other relevant quantum numbers. The dipole-dipole interaction can couple two-molecule states whose values of $m_{\rm tot} = m_A + m_B$ differ by $0,\pm 1, \pm 2$.

In the basis of Eqs.~\eqref{eq:two-mol-states}, the combined rotational and dipole-dipole Hamiltonian can be written
\begin{equation}
\begin{split}
    &H_{\rm rot} + H_{\rm dd}=\\ &\quad\begin{pmatrix}
    0 & 0 & 0 & E_{\rm dd}\\
    0 & E_{\rm rot}-E_{\rm dd} & 0 & 0\\
    0 & 0 & E_{\rm rot}+E_{\rm dd} & 0\\
    E_{\rm dd} & 0 & 0 & 2E_{\rm rot}
    \end{pmatrix}
\end{split}
\end{equation}
where $E_{\rm rot}$ is the rotational energy splitting, $E_{\rm dd} = \Lambda_{10}/4\pi\epsilon_0 |x_B-x_A|^3$ is the dipole-dipole interaction energy and the states are in the order $\{\ket{00}, \ket{\Psi^-}, \ket{\Psi^+}, \ket{11}\}$. The matrix is diagonal except for off-diagonal elements between $\ket{00}$ and $\ket{11}$.
For realistic values of $|x_B-x_A|$ in traps designed to prevent excessive overlap of the motional wavefunctions, $E_{\rm rot}\gg E_{\rm dd}$ and these elements can be safely ignored. In this case the eigenstates are well approximated by the basis of Eqs.~\eqref{eq:two-mol-states}.

\section{Effect of motion on dipole-dipole interaction}
\label{secSupp:motion}

First consider the case where the four internal states in question experience the same trap frequency $\omega_{\rm t}$, the states $\ket{0_-/1_-}$ have a trap minimum at position $x=-\delta x/2$ and the states $\ket{0_+/1_+}$ have a trap minimum at position $x=\delta x/2$. The full motional Hamiltonian is now separable from the internal part. In the subspace of Eqs.~\eqref{eq:two-mol-states} we have
\begin{equation}
\begin{split}
        H_{\rm m} &= \frac{p_A^2}{2M} + \frac{p_B^2}{2M} + \frac{1}{2}M\omega_{\rm t}^2\left(x_A+\frac{\delta x}{2}\right)^2 \\
        &\quad+ \frac{1}{2}M\omega_{\rm t}^2\left(x_B-\frac{\delta x}{2}\right)^2 +q \frac{\Lambda_{10}}{4\pi\epsilon_0|x_B-x_A|^3},
\end{split}
\label{eqSupp:ham-mot}
\end{equation}
where the factor $q=\{0,-1,1,0\}$ for states $\{\ket{00}, \ket{\Psi^-}, \ket{\Psi^+}, \ket{11}\}$.
We can reformulate this in terms of the dimensionless, relative position operators $x_{\rm cm}=\sqrt{\frac{M\omega_{\rm t}}{2\hbar}}(x_B+x_A)$ and $x_{\rm rel}=\sqrt{\frac{M\omega_{\rm t}}{2\hbar}}(x_B-x_A)$, and their conjugate momenta $p_{\rm cm}$ and $p_{\rm rel}$,
\begin{equation}
        \frac{1}{\hbar\omega_{\rm t}}H_{\rm m} = H_{\rm cm} + H_{\rm rel}
\label{eqSupp:ham-mot-red}
\end{equation}
where
\begin{subequations}
    \begin{align}
        H_{\rm cm} & = \frac{p_{\rm cm}^2}{2} + \frac{1}{2} x_{\rm cm}^2 ,\label{eqSupp:ham-cm}\\
        H_{\rm rel} & = \frac{p_{\rm rel}^2}{2} + \frac{1}{2} (x_{\rm rel}-\tilde{\delta x})^2 + q \frac{r^3}{|x_{\rm rel}|^3}.\label{eqSupp:ham-rel}
    \end{align}
\end{subequations}
Here $\tilde{\delta x}=\sqrt{\frac{M\omega_{\rm t}}{2\hbar}}\delta x$ and $r = \sqrt{\frac{M\omega_{\rm t}}{2\hbar}}(\Lambda_{10}/4\pi\epsilon_0\hbar\omega_{\rm t})^{1/3}$. The center of mass $x_{\rm cm}$ oscillates about zero with the same trap frequency as the individual particles. In the absence of the dipole-dipole interaction, the relative motion is also harmonic, oscillating around $\tilde{\delta x}$ with the same trap frequency. When the dipole-dipole interaction is included, the motion is more complicated. We can get some intuition by looking at the case where $r\ll\tilde{\delta x}$. The series expansion to second order in $(x_{\rm rel}-\tilde{\delta x})$ is
\begin{equation}
\begin{split}
    H_{\rm rel} &\simeq \frac{p_{\rm rel}^2}{2} + \frac{1}{2}(x_{\rm rel}-\tilde{\delta x})^2 \\
    &\quad+ q \frac{r^3}{\tilde{\delta x}^3}\left(1-\frac{3(x-\tilde{\delta x})}{\tilde{\delta x}} + \frac{6 (x-\tilde{\delta x})^2}{\tilde{\delta x}^2}\right) \\ &\simeq \frac{p_{\rm rel}^2}{2} + q \frac{r^3}{\tilde{\delta x}^3} 
    + \frac{1}{2}\left( 1+ \frac{12 q r^3}{\tilde{\delta x}^5}\right)\left(x-\tilde{\delta x} - \frac{3 q r^3}{\tilde{\delta x}^4} \right)^2.
\end{split}
\end{equation}
For $q r^3>0$, the trap frequency is increased and the two molecules are pushed apart slightly so that their mean separation is larger than in the absence of the dipole-dipole interaction. For $q r^3 < 0$, these effects are reversed.

\begin{figure}
    \centering
    \includegraphics{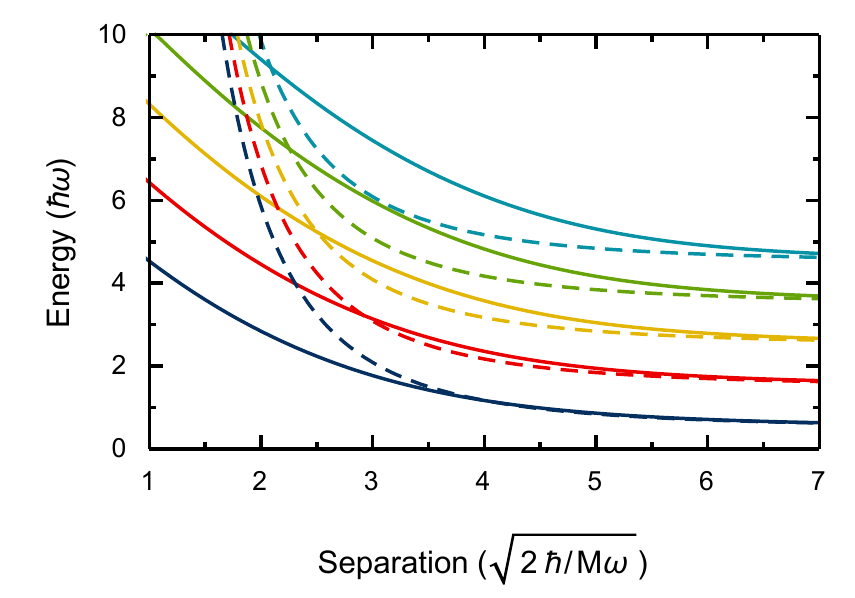}
    \caption{Energies of the first 5 relative motional states in presence of the dipole-dipole interaction for $q=1$, $r=3.5$. Dashed lines: calculations for point dipoles fixed at potential minima; solid lines: full 1D calculation.}
    \label{figSupp:eigs1D}
\end{figure}

Figure~\ref{figSupp:eigs1D} shows the energies of the first 5 eigenstates of Eq.~\eqref{eqSupp:ham-rel} calculated numerically as a function of $\tilde{\delta x}$ for $q=1$ and $r=3.5$. At large separations, the energy shifts agree well with those expected for point dipoles fixed at the potential minima (dashed lines). For intermediate separations, the shifts from the full 1D calculation are larger because the finite extent of the wavefunction means that $\langle 1/x_{\rm rel}^3\rangle > 1/\langle x_{\rm rel}\rangle^3$. This effect is larger for excited motional states where the extent of the wavefunction is larger. At small separations, the dipoles are pushed apart by their interaction and so the energy shift is reduced from the value expected from fixed point dipoles. 

\section{Hyperfine interaction and shifted potentials}

Here, we consider in more detail the complications introduced by the hyperfine interaction. As in the main text, we take the handedness of the light to be along $z$.

The hyperfine interaction couples the nuclear spin $\vec{I}$ and the total electronic angular momentum $\vec{J}$. Their sum is $\vec{F}$. Let us first consider states with well defined $F$, $J$ and $I$, and a vector Stark shift which is small compared to the hyperfine interaction so that we need only consider the diagonal matrix elements of the effective Stark shift operator. In this case, the vector Stark shift is $W_1 = \frac{1}{2 \epsilon_0 c}\alpha^{(1)} g_F m_F (\vec{C}\cdot\hat{z}) I$  where
\begin{equation}
    g_F=\frac{F(F+1)+J(J+1)-I(I+1)}{4J(J+1)F(F+1)}.
\label{eqSupp:gF}
\end{equation}

This is a useful result for molecules where the spin-rotation interaction is large compared to the hyperfine interaction. States from neighboring rotational manifolds that have the same values of $J$, $F$ and $m_F$ will have the same vector Stark shift, and the potentials for these states will be identical. However, for many molecules of interest, the hyperfine and spin-rotation interactions are similar in size so the hyperfine coupling mixes states with the same $F$ and $m_F$ but different $J$. The vector Stark shift of these mixed states is not given by Eq.~(\ref{eqSupp:gF}), but instead depends on the relative size of the hyperfine and spin-rotation coupling. As a result, in general, states in different rotational levels will have different vector Stark shifts, and potentials that are shifted relative to one another. Here we consider the effect of this shift on the dipole-dipole interaction. 

We assume the $\ket{0_\pm}$ states have potential minima at $\pm \delta x_0/2$ and the $\ket{1_\pm}$ states at $\pm \delta x_1/2$. The full Hamiltonian is now
\begin{equation}
    \begin{split}
        H =& \frac{p_A^2}{2M} + \frac{p_B^2}{2M} + \frac{1}{2}M\omega_{\rm t}^2\left(x_A+\frac{\delta x_0}{2}\right)^2\sket{0_-}{A}\sbra{0_-}{A}\\
        &\quad + \frac{1}{2}M\omega_{\rm t}^2\left(x_A+\frac{\delta x_1}{2}\right)^2\sket{1_-}{A}\sbra{1_-}{A}\\
        &\quad + \frac{1}{2}M\omega_{\rm t}^2\left(x_B-\frac{\delta x_0}{2}\right)^2\sket{0_+}{B}\sbra{0_+}{B}\\
        &\quad + \frac{1}{2}M\omega_{\rm t}^2\left(x_B-\frac{\delta x_1}{2}\right)^2\sket{1_+}{B}\sbra{1_+}{B}\\
        &\quad + \frac{\vec{d}_A\cdot\vec{d}_B-3(\vec{d}_A\cdot\hat{x})(\vec{d}_B\cdot\hat{x})}{4\pi\epsilon_0|x_B-x_A|^3}.
    \end{split}
\end{equation}
The Hamiltonian is no longer separable into motional and internal parts. We apply a unitary transformation
\begin{equation}
    \begin{split}
        U(\eta) &= \Big[T_A(-\tfrac{\eta}{2})\sket{0_-}{A}\sbra{0_-}{A} + T_A(\tfrac{\eta}{2})\sket{1_-}{A}\sbra{1_-}{A}\Big]\\
        & \otimes \Big[T_B(\tfrac{\eta}{2})\sket{0_+}{B}\sbra{0_+}{B} + T_B(-\tfrac{\eta}{2})\sket{1_+}{B}\sbra{1_+}{B}\Big]
    \end{split}
\end{equation}
where $T_{A/B}$ are the single particle translation operators such that $T_{A/B}(\eta)\ket{x_{A/B}}=\ket{x_{A/B}+\eta}$. We find
\begin{equation}
    \begin{split}
        H' &= U(\delta x_{10}) H U^\dagger(\delta x_{10})\\
        &= \frac{p_A^2}{2M} + \frac{p_B^2}{2M} + \frac{1}{2}M\omega_{\rm t}^2\left(x_A+\frac{\delta x_{\rm av}}{2}\right)^2\\
        &\quad+ \frac{1}{2}M\omega_{\rm t}^2\left(x_B-\frac{\delta x_{\rm av}}{2}\right)^2\\
         &\quad + \frac{\Lambda_{10}}{4\pi\epsilon_0}\big[D_A D_B^\dagger K(\delta x_{10}) + D_A^\dagger D_B K^\dagger(\delta x_{10})\\
         &\qquad\qquad+D_A D_B F(\delta x_{10}) + D_A^\dagger D_B^\dagger F^\dagger(\delta x_{10})\big],
    \end{split}
\end{equation}
where $\delta x_{\rm av}=\frac{1}{2}(\delta x_0+ \delta x_1)$, $\delta x_{10}=\frac{1}{2}(\delta x_0-\delta x_1)$, $D_{A} = \sket{1_-}{A}\sbra{0_-}{A}$, $D_B = \sket{1_+}{B}\sbra{0_+}{B}$ and we have defined the operators
\begin{equation}
    \begin{split}
    K(\eta) &= T_A\left(\tfrac{\eta}{2}\right)T_B\left(\tfrac{\eta}{2}\right)\frac{1}{|x_B-x_A|^3}T_A^\dagger\left(-\tfrac{\eta}{2}\right)T_B^\dagger\left(-\tfrac{\eta}{2}\right),\\
    F(\eta) &= T_A\left(\tfrac{\eta}{2}\right)T_B\left(-\tfrac{\eta}{2}\right)\frac{1}{|x_B-x_A|^3}T_A^\dagger\left(-\tfrac{\eta}{2}\right)T_B^\dagger\left(\tfrac{\eta}{2}\right).
    \end{split}
\end{equation}
We can immediately neglect, as before, the off-resonant terms in $D_A D_B$ and $D_A^\dagger D_B^\dagger$ because they couple internal states which are separated by the rotational energy.
Transforming again to the dimensionless relative position operators we have
\begin{equation}
    \begin{split}
        \tilde{H}' &= \frac{1}{\hbar\omega_{\rm t}} H' = \frac{p_{\rm cm}^2}{2} + \frac{p_{\rm rel}^2}{2} + \frac{1}{2} x_{\rm cm}^2 + \frac{1}{2} (x_{\rm rel}-\tilde{\delta x}_\mathrm{av})^2 \\&\quad +\frac{r^3}{|x_{\rm rel}|^3}\big[T_{\rm cm}(2\xi)D_A D_B^\dagger + T_{\rm cm}(-2\xi)D_A^\dagger D_B\big],
    \end{split}
    \label{eqSupp:red-mot-ham-shift}
\end{equation}
where $\xi = \sqrt{\frac{M\omega_{\rm t}}{2\hbar}}\delta x_{10}$ and $\tilde{\delta x}_\mathrm{av}=\sqrt{\frac{M\omega_{\rm t}}{2\hbar}}\delta x_\mathrm{av}$. $T_{\rm cm}$ and $T_{\rm rel}$ are the translation operators for the dimensionless center-of-mass and relative coordinates respectively and we have used $T_A(\frac{\delta x_{10}}{2}) = T_{\rm rel}(-\frac{\xi}{2})T_{\rm cm}(\frac{\xi}{2})$, $T_B(\frac{\delta x_{10}}{2}) = T_{\rm rel}(\frac{\xi}{2})T_{\rm cm}(\frac{\xi}{2})$. In Eq.~\eqref{eqSupp:red-mot-ham-shift}, the dipole-dipole interaction couples the center-of-mass and relative motions, and has an off-diagonal matrix element between the states $\ket{\Psi^+}$ and $\ket{\Psi^-}$.

We note that $\tilde{H}' = \tfrac{1}{\hbar \omega_{\rm t}}H_{\rm m} +\delta H$ where $H_{\rm m}$ is given by Eq.~(\ref{eqSupp:ham-mot-red}) and
\begin{equation}
\begin{split}
   \delta H =& \frac{r^3}{|x_{\rm rel}|^3}\big[T_{\rm cm}(2\xi)D_A D_B^\dagger+ T_{\rm cm}(-2\xi)D_A^\dagger D_B\big]\\ &- \frac{r^3}{|x_{\rm rel}|^3}\big[D_A D_B^\dagger + D_A^\dagger D_B\big].
 \end{split}
\end{equation}
Taking zeroth-order eigenstates to be those of $H_{\rm m}$, and using first-order perturbation theory, we find that the dipole-dipole energies are just multiplied by the factor $e^{-\xi^2}$. This factor is used in the calculation of the dipole-dipole interaction in the main text. It accounts for the shifted potentials irrespective of the size of the dipole-dipole interaction because the center-of-mass eigenstates of Eq.~\eqref{eqSupp:ham-cm} are unchanged by the dipole-dipole interaction.

\section{Tunneling}

States that have total angular momentum $F>1/2$ will be subject to a tensor ac Stark shift. This has several effects. One is to make the trap frequencies for different states differ from each other. For the parameters considered in this work, this effect is small. The most important effect is to give an off-diagonal term that couples together states with $\Delta m_F \le 2$. For pairs of states that are degenerate at $x=0$, this introduces an avoided crossing and provides a mechanism for molecules to tunnel from one potential to the other. 

Let $a$ be the matrix element of the tensor part of the Stark shift operator between the pair of states, and assume that this is constant across the region of interest. The single-molecule tunneling rate is
\begin{equation}
    \gamma_\mathrm{t} \simeq 2 |a| \cusbraket{\phi_-}{\phi_+}.
\end{equation}
where $\ket{\phi_-}$ and $\ket{\phi_+}$ are the motional states corresponding to the pair of internal states. If we assume harmonic oscillator ground states, this is $2 |a| e^{-\tilde{\delta x}^2/2}$. 

In the realistic example described in the main text, there is no tunneling because we choose orthogonal tweezer polarizations so that $a=0$ for all pairs of states. If instead we choose parallel polarizations, there can be tunelling between the $\ket{1_\pm}$ states with $|a|=\SI{1.4}{\mega\hertz}$. Taking the same tweezer parameters used in the main text, $\delta x$ for this pair of states is \SI{175}{\nano\meter}, and $\gamma_{\rm t}$ is \SI{2.6}{\kilo\hertz}.

\section{Photon scattering rate}

The scattering rate is dominated by the vector tweezer since it is tuned much closer to resonance than the scalar tweezer. The scattering rate from light tuned to the midpoint of the fine structure interval is dominated by scattering from these two levels and is well approximated by,
\begin{equation}
    R_\mathrm{ph} = \frac{2\Gamma\Omega^2}{3\delta_{\rm fs}^2}
\end{equation}
where $\Omega=d_{AX}\sqrt{2 I/\epsilon_0 c}/\hbar$ is the Rabi frequency and $\Gamma$ is the linewidth of the $^2\Pi$ state. For CaF we have $\Gamma=2\pi\times \SI{8.3}{\mega\hertz}$, $\delta_{\rm fs}=2\pi\times\SI{2.14}{\tera\hertz}$ and $d_{AX}=0.97\times\SI{5.95}{\debye}$ where the first factor is the Franck-Condon factor and the second is the transition dipole moment between electronic states. 

\section{Collisional loss rate}

The collisional loss rate for two molecules with wavefunctions $\psi_A(x)$ and $\psi_B(x)$ is
\begin{equation}
    R_{\rm col} = \beta \int |\psi_A(x)|^2|\psi_B(x)|^2 \mathrm{d}^3 x
\end{equation}
where $\beta$ is the two-body loss rate constant, recently measured for CaF in \SI{780}{\nano\meter} tweezer traps \cite{Cheuk2020}. For two molecules in the motional ground states of two displaced but otherwise identical potential wells, this is
\begin{equation}
    R_{\rm col} = \beta \left(\frac{m}{2 \pi\hbar}\right)^\frac{3}{2} \omega_{\rm r} \omega_{\rm a}^\frac{1}{2} e^{-\tilde{\delta x}^2}
\end{equation}
where $\omega_{\rm r}$ is the trap frequency in the radial direction and $\omega_{\rm a}$ is the trap frequency in the axial direction.

We find that, for CaF in a trap that has $\omega_{\rm r}=2\pi\times\SI{200}{\kilo\hertz}$, $\omega_{\rm a}=2\pi\times\SI{35}{\kilo\hertz}$, keeping the collisional loss rate below \SI{1}{\hertz} requires a separation of 4.7 oscillator lengths, or \SI{140}{\nano\meter}. The rate is very sensitive to separation in this region -- decreasing the separation to 3.6 harmonic oscillator lengths (or \SI{106}{\nano\meter}) increases the loss rate to \SI{100}{\hertz}.

\section{Transport}

\begin{figure}
    \centering
    \includegraphics{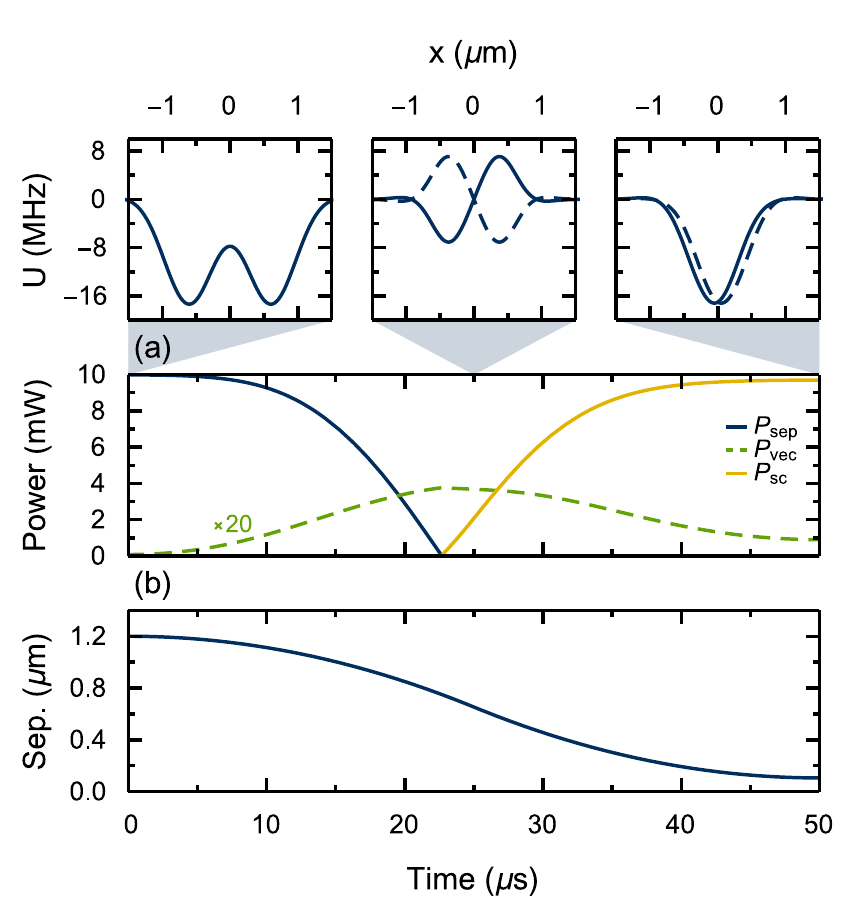}
    \caption{Trap merge sequence. (a) Intensity ramps of the different tweezers used in the sequence. (b) Separation of the potential minima as a function of time. Top row shows snapshots of potentials at three different times in the sequence.}
    \label{figSupp:transport}
\end{figure}

Figure~\ref{figSupp:transport} illustrates a simple sequence to bring a pair of CaF molecules from two separated potentials into a single, state-dependent trap ready for the fast two-qubit gate. The adiabatic transport is achieved with intensity ramps of four spatially-fixed tweezer traps as shown in Fig.~\mref{figSupp:transport}{(a)}. The top row of the figure shows the potentials at three points during the merger. The molecules begin in two \SI{780}{\nano\meter} tweezers focused at $x=\pm\SI{0.6}{\micro\meter}$, each having power $P_{\rm sep}$. A \SI{604.966}{\nano\meter} tweezer and another \SI{780}{\nano\meter} tweezer are focused at $x=0$ and have powers $P_{\rm vec}$ and $P_{\rm sc}$ respectively. $P_{\rm sep}$ is ramped down from its initial value of $\SI{10}{\milli\watt}$ while $P_{\rm vec}$ is ramped up (note that $P_{\rm vec}$ is shown multiplied by 20 to appear on same scale). Finally $P_{\rm sc}$ is ramped up to squeeze the two molecules together. Figure~\mref{figSupp:transport}{(b)} shows the separation of the trap minima as a function of time. The chosen power ramps keep the trap frequency fixed throughout. The mean number of photons scattered during this transport sequence is \num{1.6e-2} and the probability of motional excitation is smaller still. More sophisticated non-adiabatic sequences will allow for faster transport and fewer scattered photons.
